\documentstyle[12pt]{article}

\global\arraycolsep=2pt

\begin{document}

\begin{titlepage}

\begin{flushright}
CERN-TH.7113/93\\
hep-ph/9312311
\end{flushright}

\vspace{0.3cm}

\begin{center}
\Large\bf Analysis of the Photon Spectrum\\
in Inclusive $B\to X_s\,\gamma$ Decays
\end{center}

\vspace{0.8cm}

\begin{center}
Matthias Neubert\\
{\sl Theory Division, CERN, CH-1211 Geneva 23, Switzerland}
\end{center}

\vspace{0.8cm}

\begin{abstract}
Using a combination of the operator product and heavy quark
expansions, we resum the leading nonperturbative contributions to the
inclusive photon spectrum in $B\to X_s\,\gamma$ decays. The shape of
the spectrum is determined by a universal structure function, which
describes the distribution of the light-cone momentum of the
$b$-quark inside the $B$-meson. The moments of this function are
proportional to forward matrix elements of higher-dimension
operators. As a by-product, we obtain the bound $\lambda_1<0$ for one
of the parameters of the heavy quark effective theory. The integral
over the $B\to X_s\,\gamma$ structure function is related to the
shape function that governs the fall-off of the lepton spectrum close
to the endpoint in $B\to X_u\,\ell\,\bar\nu$ decays. A measurement of
the photon spectrum in rare $B$-decays can therefore help to obtain a
model-independent determination of $V_{ub}$.
\end{abstract}

\centerline{(submitted to Physical Review D)}
\bigskip\bigskip

\noindent
December 1993

\end{titlepage}

\section{Introduction}

The rare decays of $B$-mesons are of great importance, since they are
sensitive probes of new physics beyond the standard model (see, e.g.,
Refs.~\cite{Barb,Rand}). Such processes are induced by the exchange
of heavy particles, which manifests itself at low energies in the
appearance of local operators multiplied by small coefficient
functions. These coefficients depend on the masses and quantum
numbers of the heavy particles. In the standard model, rare decays of
the type $b\to s\,\gamma$ are induced by penguin diagrams with
virtual top or charm quark exchange. Recently, the first such decay
mode, $B\to K^*\gamma$, has been observed by the CLEO collaboration.
The reported branching ratio is $(4.5\pm 1.9\pm 0.9)\%$ \cite{CLEO}.
The interpretation of this result is difficult, however, due to the
lack of reliable theoretical methods to calculate the low-energy
hadronic matrix element of the penguin operator between meson states.
Existing estimates of this matrix element rely on quark models
\cite{Alto,Desh} or QCD sum rules \cite{Domi,Alie,Ball} and are thus
model-dependent.

For several reasons, one expects that the theoretical analysis is
more reliable for inclusive $B\to X_s\,\gamma$ decays, where one
sums over all possible final states containing a strange particle.
Assuming quark-hadron duality, the inclusive decay rate was
traditionally calculated using the free quark decay model and
including short-distance corrections from virtual and real gluons
\cite{Shif,Camp,Bert,Grin,Grig,Cell,Misi,Ciuc,Ali}. Recently,
however, it has been observed that inclusive decays of hadrons
containing a heavy quark $Q$ allow for a systematic expansion in
powers of $\Lambda/m_Q$, where $\Lambda$ is a characteristic low
energy scale of the strong interactions
\cite{Chay,Bigi,Blok,MaWe,Adam,Thom}. The parton model emerges as the
leading term in this QCD-based expansion, and the nonperturbative
corrections to it are suppressed by a factor $\Lambda^2/m_Q^2$. The
fact that there are no first-order power corrections relies on a
particular definition of $m_Q$, which is provided in a natural way by
requiring that there be no residual mass term for the heavy quark in
the heavy quark effective theory \cite{AMM,SR3}. This definition is
unique and can be regarded as a nonperturbative generalization of the
concept of a pole mass.

The availability of a systematic expansion of the inclusive $B\to
X_s\,\gamma$ decay rate raises the hope for a better understanding of
rare decays, which is necessary to increase the sensitivity to new
physics. For practical reasons, however, it is not sufficient to have
a reliable calculation of the total decay rate. In fact, the
distribution of the photon energy will be affected by various
experimental cuts, and it is thus the spectrum ${\rm d}\Gamma/{\rm
d}E_\gamma$ that needs to be calculated. In the free quark decay
model, the photon in $b\to s\,\gamma$ decays is monochromatic.
Corrections to this simple picture arise from two sources: Real gluon
emission produces three-body final states, leading to a continuous
energy spectrum. These effects have been calculated in perturbation
theory \cite{Ali}; they will not be discussed here. In addition,
bound-state corrections in the initial state, in particular the
``Fermi motion'' of the $b$-quark, lead to a dispersion of the
spectrum. These nonperturbative effects can Doppler-shift the
spectrum above the parton model endpoint. So far, such effects have
been estimated \cite{Greub} using the phenomenological model of
Altarelli et al.~\cite{ACM}.

In this paper, we present a more rigorous treatment of bound-state
corrections to the free quark decay model. In particular, we show
that QCD provides a natural framework to account for the ``Fermi
motion''. Extending our previous analysis of the endpoint region of
the lepton spectrum in $B\to X_u\,\ell\,\bar\nu$ decays \cite{shape},
we resum the leading nonperturbative contributions to the photon
spectrum in $B\to X_s\,\gamma$ decays to all orders in the $1/m_b$
expansion. We show that the spectrum is determined by a fundamental
structure function, which describes the light-cone residual momentum
distribution of the heavy quark inside the $B$-meson. Quite
remarkably, an integral over this structure function is related to
the shape function that governs the endpoint region of the lepton
spectrum in $B\to X_u\,\ell\,\bar\nu$ decays. In Sect.~\ref{sec:2},
we generalize the results of Refs.~\cite{Bigi,Adam} and construct the
operator product expansion for the photon spectrum in $B\to
X_s\,\gamma$ decays, including the leading nonperturbative
corrections. In Sect.~\ref{sec:3}, we perform a resummation of the
most singular terms in the photon spectrum to all orders in $1/m_b$.
The moments of the spectrum are related to forward matrix elements of
local, higher-dimension operators in the heavy quark effective
theory. We show that the characteristic width of the spectrum is
determined by the kinetic energy of the $b$-quark inside the
$B$-meson. From this relation, we derive the bound $\lambda_1<0$ for
one of the fundamental parameters of the effective theory. In
Sect.~\ref{sec:4}, we illustrate our results using a toy model, which
is a simplified version of the approach of Ref.~\cite{ACM}. In
Sect.~\ref{sec:5}, we derive the relation between the photon spectrum
and the $b$-quark structure function of the $B$-meson, and we relate
the Fourier transform of the spectrum to the forward matrix element
of a gauge-invariant, bilocal operator. Section~\ref{sec:6} is
devoted to a discussion of the connection between rare $B$-decays and
inclusive semileptonic decays. We show that the shape function
\cite{shape}, which describes the fall-off of the lepton spectrum
close to the endpoint in $B\to X_u\,\ell\,\bar\nu$ decays, is given
by an integral over the photon spectrum in rare $B\to X_s\,\gamma$
decays, up to corrections of order $1/m_b$. This connection may help
to obtain a model-independent determination of $V_{ub}$. In
Sect.~\ref{sec:7}, we summarize our results and give some
conclusions.

\section{Operator Product Expansion}
\label{sec:2}

In this section, we discuss the application of the operator product
expansion to the inclusive rare decays $B\to X_s\,\gamma$. We derive
expressions for the photon spectrum and the total decay rate, to
order $1/m_b^2$ in the heavy quark expansion and to leading
logarithmic order in $\alpha_s(m_b)$. In the limit where the mass of
the strange quark is neglected, the total decay rate and the average
photon energy have been calculated in Refs.~\cite{Bigi,Adam}. We will
generalize the results presented there.

In leading logarithmic approximation, the rare decays of interest are
mediated by an effective Hamiltonian containing a local penguin
operator,
\begin{equation}\label{Heff}
   {\cal{H}}_{\rm eff} = - {4 G_F\over\sqrt{2}}\,V_{tb} V_{ts}^*\,
   c_7(m_b)\,{e\over 16\pi^2}\,\bar s\,\sigma^{\mu\nu}\,
   (m_b\,P_R + m_s\,P_L)\,b\,F_{\mu\nu} \,,
\end{equation}
where $P_L=\frac{1}{2}(1-\gamma_5)$ and $P_R=\frac{1}{2}(1+\gamma_5)$
are left- and right-handed projection operators, and $F_{\mu\nu}$ is
the electromagnetic field strength tensor. The Wilson coefficient
$c_7(m_b)$ describes the evolution from high-energy scales $\mu\sim
m_t$ or $m_W$ to low-energy scales $\mu\sim m_b$
\cite{Shif,Camp,Bert,Grin,Grig,Cell,Misi,Ciuc,Ali}. It is sensitive
to the mass of the top quark and, more generally, to any kind of new
physics beyond the standard model.

In terms of the effective Hamiltonian, the inclusive differential
decay rate is given by
\begin{equation}
   {\rm d}\Gamma = {{\rm d}^3\vec p_\gamma\over(2\pi)^3\,2 E_\gamma}
   \sum_{X_s,{\rm pol}} (2\pi)^4\,\delta^4(m_B v-p_\gamma-p_X)\,
   \Big| \langle X_s(p_X)\,\gamma(p_\gamma)|\,{\cal{H}}_{\rm eff}\,
   |B(v)\rangle \Big|^2 \,,
\end{equation}
where we sum over the two transverse polarization states of the
photon. We use a mass-independent normalization of states such that
\begin{equation}
   \langle B(v)|B(v)\rangle = v^0\,(2\pi)^3\,\delta^3(\vec 0) \,,
\end{equation}
where $v$ is the four-velocity of the $B$-meson. The decay rate can
be written in terms of the imaginary part of a correlator of two
local currents, which contains all dependence on hadronic dynamics.
We define
\begin{equation}
   T(v,p_\gamma) = -i\int{\rm d}x\,e^{i(m_b v-p_\gamma)\cdot x}\,
   \langle B(v)|\,{\rm T}\,\big\{ J^\mu(x), J_\mu^\dagger(0)
   \big\}\,|B(v)\rangle \,,
\end{equation}
where
\begin{equation}\label{Jdef}
   J^\mu(x) = \bar b_v(x)\,[\gamma^\mu,\rlap{\,/}p_\gamma]\,
   (m_b\,P_L + m_s\,P_R)\,s(x) \,;\quad
   b_v(x) = e^{i m_b v\cdot x}\,b(x) \,.
\end{equation}
This leads to
\begin{equation}\label{dGam}
   {\rm d}\Gamma = {{\rm d}^3\vec p_\gamma\over(2\pi)^3\,2E_\gamma}
   \,{G_F^2\,\alpha\over 8\pi^3}\,|\,V_{tb} V_{ts}^*|^2\,
   |c_7(m_b)|^2\,{\rm Im}\,T(v,p_\gamma) \,.
\end{equation}
Note that the Fourier components of the rescaled heavy quark field
$b_v(x)$ in (\ref{Jdef}) contain the ``residual'' momentum $k=p_b-m_b
v$. In contrast with the full heavy quark momentum $p_b$, the
residual
momentum is of order $\Lambda_{\rm QCD}$. It is then appropriate to
construct an expansion in powers of $k/m_b$.

To this end, it is convenient to introduce dimensionless variables
\begin{equation}
   \hat p = {p_\gamma\over m_b} \,,\qquad
   \hat m = \sqrt{\rho}= {m_s\over m_b} \,,\qquad
   y = 2 v\cdot\hat p = {2 E_\gamma\over m_b} \,,
\end{equation}
where $E_\gamma$ denotes the photon energy in the rest frame of the
$B$-meson. Since $v^2=1$ and $p_\gamma^2=0$, the function
$T(v,p_\gamma)$ only depends on the kinematic variable $y$. The
correlator $T(y)$ is analytic in the complex $y$-plane, with
discontinuities on the real axis. The physical region corresponding
to the decay $B\to X_s\,\gamma$ is
\begin{equation}\label{kine}
   0\le y\le {m_B\over m_b}\,\bigg( 1 - {m_{K^*}^2\over m_B^2}
   \bigg) \simeq 1.09 \,,
\end{equation}
where we have assumed $m_b\simeq 4.7$ GeV for the purpose of
illustration. In addition, there is a cut starting at $y\simeq 4.16$
corresponding to the process $\gamma+B\to X_{sbb}$, where $X_{sbb}$
contains two $b$-quarks and an $s$-quark. This unphysical cut is
separated from the physical one by a large energy gap $\Delta
E_\gamma = m_B\,(1+m_{K^*}/m_B)^2\simeq 7.2$ GeV. Because of this
analytic structure, phase-space integrals of $T(y)$ with smooth
weight functions can be deformed from the physical region into
contour integrals far away from the physical singularities. It is
then possible to construct an operator product expansion of the
correlator in terms of local operators ${\cal{O}}_i$, which contain
the $b$-quark fields and have dimension $d\ge 3$
\cite{Chay,Bigi,Blok,MaWe,Adam,Thom}. These operators are multiplied
by short-distance coefficient functions $C_i(y)$, which can be
computed in perturbation theory using free quark states. The leading
contributions in $\alpha_s$ come from a tree diagram with an
intermediate $s$-quark carrying the momentum $(p_b-p_\gamma) = (m_b
v+k-p_\gamma)$. The operator product expansion is obtained by
replacing the residual momentum $k$ by a covariant derivative $i D$.
This gives the propagator in the background field of the light
degrees of freedom in the decaying $B$-meson. Next, one expands the
propagator in powers of $iD/m_b$. This gives
\begin{equation}\label{propa}
   {1\over m_b(\rlap/v-\rlap{\,/}\hat p-\hat m+i\epsilon)
   + i\rlap{\,/}D}
   = {\rlap/v-\rlap{\,/}\hat p+\hat m\over m_b\,\Delta}\,
   \sum_{n=0}^\infty\,\Bigg[ -i\rlap{\,/}D\,
   {(\rlap/v-\rlap{\,/}\hat p+\hat m)\over m_b\,\Delta} \Bigg]^n \,,
\end{equation}
where
\begin{equation}\label{Deldef}
   \Delta = (v-\hat p)^2 - \hat m^2 + i\epsilon
   = 1-y-\rho+i\epsilon \,.
\end{equation}

Once the Wilson coefficients have been determined, one has to
evaluate the forward matrix elements of the local operators
${\cal{O}}_i$ between $B$-meson states. The leading operators have
dimension three and can be related to matrix elements of vector and
axial vector currents. The vector current matrix element is
normalized by current conservation,
\begin{equation}
   \langle B(v)|\,\bar b_v\,\gamma^\mu\,b_v\,|B(v)\rangle
   = v^\mu \,,
\end{equation}
whereas the matrix element of the axial vector current vanishes by
parity invariance. These leading-order contributions reproduce the
parton model. The nonperturbative corrections to it are described by
the matrix elements of higher-dimension operators. They can be
evaluated using the powerful formalism of the heavy quark effective
theory \cite{habil}. The heavy quark field $b_v$ is split into
``large'' and ``small'' two-component spinors $h_v=\frac{1}{2}
(1+\rlap/v)\,b_v$ and $H_v=\frac{1}{2}(1-\rlap/v)\,b_v$, and the
field $H_v$ is integrated out to obtain an effective Lagrangian
\cite{Eich,Geor,Mann}. Any operator in the full theory has an
expansion in terms of operators in the effective theory, which only
contain the field $h_v$. Using techniques \cite{MaWe,Adam,FaNe} that
are standard by now, we obtain the relations
\begin{eqnarray}
   \langle B(v)|\,\bar b_v\,\Gamma\,i D^\mu\,b_v\,|B(v)\rangle
   &=& {\lambda_1+3\lambda_2\over 12 m_b}\,\,{\rm Tr}\,\Big\{
    \Gamma\,(\gamma^\mu + v^\mu - 5 v^\mu P_+) \Big\}
    \,, \nonumber\\
   \langle B(v)|\,\bar b_v\,\Gamma\,i D^\mu\,i D^\nu\,b_v\,
   |B(v)\rangle &=& {\lambda_1\over 6}\,(g^{\mu\nu}
    - v^\mu v^\nu)\,\,{\rm Tr}\,\Big\{ \Gamma\,P_+ \Big\}
    \nonumber\\
   &&\mbox{}+ {\lambda_2\over 4}\,\,{\rm Tr}\,\Big\{ \Gamma\,P_+\,
    i\sigma^{\mu\nu} P_+ \Big\} \,,
\end{eqnarray}
where $P_+=\frac{1}{2}(1+\rlap/v)$, and $\Gamma$ denotes an arbitrary
combination of Dirac matrices. The low-energy parameters $\lambda_1$
and $\lambda_2$ are related to the kinetic energy $K_b$ of the heavy
quark inside the $B$-meson, and to the hyperfine mass splitting
between $B$- and $B^*$-mesons. They are defined as \cite{FaNe}
\begin{equation}\label{lamdef}
   K_b = -{\lambda_1\over 2 m_b} \,,\qquad
   m_{B^*}^2 - m_B^2 = 4\lambda_2 \,.
\end{equation}

Using these results, we find
\begin{eqnarray}
   T(y) = - {4 m_b^3\,(1+\rho)\,y^2\over\Delta}\,\bigg\{ 1
   &-& {\lambda_1\over 6 m_b^2}\,\bigg( {2 y^2\over\Delta^2}
    + {7 y\over\Delta} + 5 \bigg) \nonumber\\
   &-& {\lambda_2\over 2 m_b^2}\,\bigg( {5 y-8\over\Delta}
    + 5 \bigg) + {\cal{O}}(m_b^{-3}) \bigg\} \,,
\end{eqnarray}
with $\Delta$ as given in (\ref{Deldef}). In the limit $\rho=0$, this
agrees with Ref.~\cite{Adam}. For the inclusive photon spectrum, we
obtain from (\ref{dGam})
\begin{equation}\label{Gamres}
   {{\rm d}\Gamma\over{\rm d}y}
   = {G_F^2\,\alpha\,m_b^5\over 32\pi^4}\,|\,V_{tb} V_{ts}^*|^2\,
   |c_7(m_b)|^2\,(1+\rho) (1-\rho)^3\,\eta_b\,s(y,\rho) \,,
\end{equation}
where
\begin{equation}
   \eta_b = 1 + {\lambda_1 - 9\kappa\lambda_2\over 2 m_b^2}
   + {\cal{O}}(m_b^{-3}) \,;\qquad
   \kappa = {3+5\rho\over 3-3\rho} \,.
\end{equation}
The spectral function $s(y,\rho)$ is given by
\begin{eqnarray}\label{sfun}
   s(y,\rho) &=& \delta(1-y-\rho) - (1-\rho)\,
    {\lambda_1 + 3\kappa\lambda_2\over 2 m_b^2}\,\delta'(1-y-\rho)
    \nonumber\\
   &&\mbox{}- (1-\rho)^2\,{\lambda_1\over 6 m_b^2}\,
    \delta''(1-y-\rho) + {\cal{O}}(m_b^{-3}) \,.
\end{eqnarray}
The interpretation of the singular structure of this function is the
main subject of this paper. Recall that the operator product
expansion, which led to (\ref{Gamres}), was only justified when the
spectrum is integrated with a smooth weight function. Hence, one
should understand the singular expression (\ref{sfun}) in the sense
of distributions. Integrated quantities such as the total decay rate
and the average photon energy obey a well-defined $1/m_b$ expansion.
We find
\begin{eqnarray}\label{Gamma}
   \Gamma &=& {G_F^2\,\alpha\,m_b^5\over 32\pi^4}\,
    |\,V_{tb} V_{ts}^*|^2\,|c_7(m_b)|^2\,(1+\rho) (1-\rho)^3\,
    \eta_b \,, \nonumber\\
   && \\
   \langle\,y\,\rangle &=& {2\over m_b}\,\langle\,E_\gamma\,\rangle
    = (1-\rho)\,\bigg\{ 1
    - {\lambda_1 + 3\kappa\lambda_2\over 2 m_b^2}
    + {\cal{O}}(m_b^{-3}) \bigg\} \,. \nonumber
\end{eqnarray}
In the limit $\rho=0$, we confirm the results of
Refs.~\cite{Bigi,Adam}.

In order to obtain an estimate of the magnitude of the
nonperturbative corrections, we use the quark masses $m_b=4.7$ GeV
and $m_s=0.2$ GeV, corresponding to $\rho\simeq 2\times 10^{-3}$.
{}From the observed value of the $B^*$-$B$ mass splitting, one
obtains $\lambda_2\simeq 0.12$ GeV$^2$. The parameter $\lambda_1$ is
not directly related to an observable. The field-theory analogue of
the virial theorem relates the kinetic energy of a heavy quark inside
a hadron (and thus $\lambda_1$) to a matrix element of the gluon
field strength tensor \cite{virial}. This theorem makes explicit an
``intrinsic smallness'' of $\lambda_1$, which was not taken into
account in existing QCD sum rule calculations of this parameter
\cite{Subl,Elet,BaBr}. As a consequence, we expect that
$(-\lambda_1)$ is smaller than predicted in these analyses. Here we
shall use the range $-\lambda_1=0.1$--0.3 GeV$^2$. We then obtain
$\eta_b\simeq 0.97$, corresponding to a 3\% decrease of the parton
model decay rate. The correction to the average photon energy is
below 1\%.

\section{Resummation of the Most Singular Terms}
\label{sec:3}

The numerical analysis presented above shows that the nonperturbative
corrections to integrated quantities such as the total decay rate are
very small. Were it just for these corrections, one could say that
the most important result of the formalism presented in the previous
section would be to give a theoretical justification for the parton
model. However, in this section we will show that much more
interesting information can be extracted from this analysis. This
information is encoded in the coefficients of the singular terms in
the spectral function $s(y,\rho)$.

The singularities at $y=1-\rho$ in (\ref{sfun}) arise from the fact
that the operator product expansion becomes singular when the
$s$-quark propagator is almost on-shell \cite{Adam}. Although it is
legitimate to evaluate phase-space averages of smooth functions using
the singular theoretical expression, one cannot trust the shape of
the spectrum as given in (\ref{sfun}). The true spectrum will be
different. A similar situation is encountered in inclusive
semileptonic $B\to X_u\,\ell\,\bar\nu$ decays, where the lepton
spectrum obtained from the operator product expansion becomes
singular when the lepton energy approaches the parton model endpoint
\cite{Bigi,Blok,MaWe}. In Ref.~\cite{shape}, we have suggested that
the deviation of the true lepton spectrum from the parton model
prediction could be described by a shape function $S(y)$, which has a
support only in a small region close to the endpoint. The
singularities in the theoretical spectrum can be identified with the
first few terms in a moment expansion of the shape function.

We can adopt a similar point of view in the case of $B\to X_s\,
\gamma$ decays. Let us identify the function $s(y,\rho)$ in
(\ref{Gamres}) with the {\sl physical\/} photon spectrum subject to
the normalization condition
\begin{equation}
   \int\limits_0^\infty\!{\rm d}y\,s(y,\rho) = 1 \,,
\end{equation}
so that the total rate is given by the first equation in
(\ref{Gamma}). We will show below that the characteristic width
$\sigma_y$ of the spectrum is proportional to $1/m_b$. When
integrated with a smooth function that is slowly varying on scales of
order $1/m_b$, the spectral function $s(y,\rho)$ can be replaced by a
singular expansion \cite{shape}:
\begin{equation}\label{sdef}
   s(y,\rho) = \sum_{n=0}^\infty\,{M_n(\rho)\over n!}\,
   \delta^{(n)}(1-y-\rho) \,.
\end{equation}
The moments $M_n(\rho)$ are defined as
\begin{equation}\label{Mndef}
   M_n(\rho) = \int\limits_0^\infty\!{\rm d}y\,(y-1+\rho)^n\,
   s(y,\rho) \,.
\end{equation}
By definition, $M_0(\rho)=1$. We can now identify the singular
expression in (\ref{sfun}) with the first few terms in this expansion
and relate the moments $M_n(\rho)$ to nonperturbative parameters.
Neglecting terms of order $1/m_b^3$, we obtain
\begin{eqnarray}\label{M1M2}
   M_1(\rho) &=& -(1-\rho)\,
    {\lambda_1 + 3\kappa\lambda_2\over 2 m_b^2} \,, \nonumber\\
   M_2(\rho) &=& -(1-\rho)^2\,{\lambda_1\over 3 m_b^2}
    \equiv \sigma_y^2 \,.
\end{eqnarray}
{}From dimensional analysis, it follows that the moments obey an
expansion of the form
\begin{equation}\label{Mnexp}
   M_n(\rho) = {a_n(\rho)\over m_b^n} + {b_n(\rho)\over m_b^{n+1}}
   + \ldots,
\end{equation}
with coefficients $a_n(\rho)$ and $b_n(\rho)$ that are independent of
$m_b$ (up to logarithms arising from radiative corrections). Hence,
the QCD prediction that $M_1(\rho)$ is of order $1/m_b^2$ indicates a
non-trivial cancellation: The shift in the average value of $y$ due
to
bound-state effects is of order $1/m_b^2$, corresponding to a shift
of order $\Lambda_{\rm QCD}^2/m_b$ in the average photon energy.
Naively, one would expect this shift to be of order $\Lambda_{\rm
QCD}$.

The second moment is a measure of the width of the photon spectrum.
As stated above, we find that $\sigma_y\propto 1/m_b$. In
Ref.~\cite{shape}, we showed that the quantity $\sigma_y$ evaluated
for $\rho=0$ describes the width of the endpoint region of the lepton
spectrum in $B\to X_u\,\ell\,\bar\nu$ decays. The connection between
these two cases will be discussed in more detail in
Sect.~\ref{sec:6}.
Note that the width is determined by the parameter $(-\lambda_1)$,
which is proportional to the kinetic energy of the $b$-quark inside
the $B$-meson [see~(\ref{lamdef})]. Since, by definition, the second
moment is positive, we obtain the upper bound $\lambda_1<0$. Although
this result is certainly not surprising, it is not trivial, since
operator renormalizations could spoil the positive definiteness of
the kinetic energy operator that defines $(-\lambda_1)$ \cite{BiUr}.

For an understanding of the properties of the spectrum close to the
endpoint, it is not sufficient to truncate the $1/m_b$ expansion at
order $1/m_b^2$ \cite{shape}. What is relevant are the rescaled
moments
\begin{equation}\label{Mnresc}
   \int\limits_0^\infty\!{\rm d}E_\gamma\,
   \Big[ 2 E_\gamma-m_b\,(1-\rho)\Big]^n\,s(E_\gamma,\rho)
   = a_n(\rho) + {b_n(\rho)\over m_b} + \ldots,
\end{equation}
where $s(E_\gamma,\rho)\equiv (2/m_b)\,s(y,\rho)$, such that
$\int_0^\infty\!{\rm d}E_\gamma\,s(E_\gamma,\rho)=1$. These moments
are all equally important in the limit $m_b\to\infty$. Hence, it is
necessary to resum the operator product expansion. However, with the
exception of the first moment, for which the coefficient $a_1(\rho)$
vanishes, one may argue that it is a good approximation to keep the
leading coefficient $a_n(\rho)$ for each moment. The corrections
involving $b_n(\rho)$ change the moments by small amounts.

The aim is thus to construct a partial resummation of the operator
product expansion, in which one keeps the leading term in each moment
but neglects $1/m_b$ corrections. A crucial observation is that, at
any order in the $1/m_b$ expansion, the coefficient $a_n(\rho)$
receives contributions only from the most singular terms in the
theoretical expression for $s(y,\rho)$. For instance, at order
$1/m_b^2$ the coefficient of the $\delta''$-function in (\ref{sfun})
determines $a_2(\rho)$; the coefficient of the $\delta'$-function,
however, determines $b_1(\rho)$. A resummation of the most singular
terms can be constructed by using the following alternative way of
writing the $s$-quark propagator in (\ref{propa}):
\begin{eqnarray}
   {1\over m_b(\rlap/v-\rlap{\,/}\hat p-\hat m+i\epsilon)
    + i\rlap{\,/}D} = {\rlap/v-\rlap{\,/}\hat p+\hat m\over
    m_b \Delta + 2(v-\hat p)\cdot iD} \nonumber\\
   \times \sum_{n=0}^\infty\,
    \Bigg[ (\rlap/v-\rlap{\,/}\hat p-\hat m)\,i\rlap{\,/}D\,
    {1\over m_b \Delta + 2(v-\hat p)\cdot iD} \Bigg]^n \,.
\end{eqnarray}
Note that all terms but the first one are multiplied by a factor
$\Delta=(\rlap/v-\rlap{\,/}\hat p+\hat m)(\rlap/v-\rlap{\,/}\hat p
-\hat m)$. Since $\Delta$ vanishes at the endpoint, it follows that
the first term is more singular than the other ones. Hence, the
leading singularities can be resummed by using the replacement
\begin{equation}
   {1\over m_b(\rlap/v-\rlap{\,/}\hat p-\hat m+i\epsilon)
   + i\rlap{\,/}D} \to {\rlap/v-\rlap{\,/}\hat p+\hat m \over
   m_b \Delta + 2(v-\hat p)\cdot iD}
\end{equation}
for the $s$-quark propagator. The imaginary part of this expression
is given by a $\delta$-function, and it is straightforward to find
that
\begin{equation}\label{sresult}
   s(y,\rho) = \bigg\langle \delta\bigg[ 1-y-\rho + {2\over m_b}\,
   (v-\hat p)\cdot iD\bigg] \bigg\rangle
   +~\hbox{less singular terms,}
\end{equation}
where we define the expectation value of an operator $O$ as
\begin{equation}
   \langle\,O\,\rangle =
   {\langle B(v)|\,\bar h_v\,O\,h_v\,|B(v)\rangle \over
    \langle B(v)|\,\bar h_v\,h_v\,|B(v)\rangle} \,.
\end{equation}
Here, $h_v$ is the velocity-dependent heavy quark field in the heavy
quark effective theory \cite{Geor,Mann}, and the states are the
eigenstates of the corresponding effective Lagrangian.\footnote{It is
sufficient to evaluate the matrix elements in the effective theory,
since we are not interested in the higher-order corrections in
(\ref{Mnexp}).}
Equation~(\ref{sresult}) is a formal definition of the spectral
function, which is valid to all orders in the $1/m_b$ expansion. The
``less singular terms'' omitted here do not contribute to the
leading-order coefficients in the expansion of the moments. Expanding
our result in powers of $1/m_b$, and comparing with (\ref{sdef}), we
find
\begin{equation}\label{anres}
   a_n(\rho) = \Big\langle\,\Big[ 2 (v-\hat p)\cdot i D\Big]^n\,
   \Big\rangle\Big|_{y=1-\rho} \,.
\end{equation}

In order to extract further information from this relation, it is
necessary to investigate the structure of forward matrix elements of
higher-dimension operators in the heavy quark effective theory. Using
the equation of motion, $i v\cdot D\,h_v=0$, one can show that
\cite{shape}
\begin{eqnarray}\label{Andef}
   \langle\,i D^\mu\,\rangle &=& 0 \,, \nonumber\\
   \phantom{ \Big[ }
   \langle\,i D^\mu\,i D^\nu\,\rangle &=& A_2\,
    (v^\mu v^\nu - g^{\mu\nu}) \,, \nonumber\\
   \phantom{ \Big[ }
   \langle\,i D^\mu\,i D^\nu\,i D^\alpha\,\rangle
   &=& A_3\,(v^\mu v^\alpha - g^{\mu\alpha})\,v^\nu \,,
    \nonumber\\
   \phantom{ \Big[ }
   \langle\,i D^{\mu_1}\ldots\,iD^{\mu_n} \rangle
   &=& A_n\,v^{\mu_1}\ldots\,v^{\mu_n}
    +~\hbox{terms with $g^{\mu_i\mu_j}$,}
\end{eqnarray}
where $A_2=-\lambda_1/3$ \cite{FaNe}. For $n\ge 4$, the matrix
elements can no longer be parameterized by a single parameter $A_n$.
However, terms involving the metric tensor give only small
contributions, of order $\rho$, to the coefficients $a_n(\rho)$ in
(\ref{anres}). Hence, we obtain $a_0(\rho)=1$, $a_1(\rho)=0$, as well
as
\begin{eqnarray}\label{a2a3}
   a_2(\rho) &=& -(1-\rho)^2\,{\lambda_1\over 3} \,, \nonumber\\
   a_3(\rho) &=& (1-\rho)^2 (1+\rho)\,A_3 \,, \nonumber\\
   \phantom{ \bigg[ }
   a_n(\rho) &=& A_n + {\cal{O}}(\rho) \,;\quad n\ge 4 \,.
\end{eqnarray}

Let us summarize the main results of this section: Apart from
radiative corrections, the inclusive photon spectrum in $B\to
X_s\,\gamma$ decays can be described by a spectral function
$s(y,\rho)$, which is a genuinely nonperturbative form factor that
accounts for bound-state effects in the decaying meson. The moments
of this function obey a well-defined $1/m_b$ expansion. The leading
terms in this expansion are related to forward matrix elements of
higher-dimension operators in the heavy quark effective theory. In
the limit $\rho=0$, these matrix elements are described by a set of
fundamental parameters $A_n$. Since the even moments of the spectral
function are positive, it follows that $A_{2n}>0$. In particular,
this gives the upper bound $\lambda_1<0$.

\section{A Toy Model}
\label{sec:4}

Before we discuss in more detail the physics of the results derived
in the previous section, we find it instructive to illustrate them in
the framework of a toy model, which is a simplified version of the
phenomenological approach of Altarelli et al.\ (ACM) \cite{ACM}. In
the ACM model, the validity of the parton model is assumed, and
bound-state effects are incorporated by assigning a momentum
distribution $\phi(|\vec p_b|)$ to the heavy quark.\footnote{In
addition, the heavy quark mass is treated as a momentum-dependent
parameter $m_b(|\vec p_b|)$. For simplicity, we shall not consider
this aspect of the model.} It is then appropriate to replace the
covariant derivative by the spatial components of the heavy quark
momentum $\vec p_b$. The gluon field and the time-component of the
covariant derivative are neglected. Accordingly, in the rest frame of
the $B$-meson, one makes the replacement
\begin{equation}
   2(v-\hat p)\cdot i D\Big|_{y=1-\rho}
   \to (1-\rho)\,p_{\|} \,,
\end{equation}
where $p_{\|}=\vec p_b\cdot p_\gamma/|\vec p_\gamma|$ denotes the
component of the $b$-quark momentum in the photon direction. The
matrix elements in (\ref{sresult}) and (\ref{anres}) are replaced by
integrals over the momentum distribution of the heavy quark. In the
ACM model, one assumes a Gaussian momentum distribution,
\begin{equation}
   \phi(|\vec p_b|) = {4\over\sqrt{\pi}\,p_{\rm F}^3}\,
   \exp\bigg(-{|\vec p_b|^2\over p_{\rm F}^2}\bigg) \,,
\end{equation}
where $p_{\rm F}$ is the Fermi momentum. It is straightforward to
calculate the hadronic matrix elements $a_n(\rho)$ in this toy model.
We find
\begin{equation}\label{antoy}
   a_n^{\rm toy}(\rho) = (1-\rho)^n\,\langle\,p_{\|}^n\,\rangle
   = \cases{ (1-\rho)^n\,\displaystyle{(n-1)!!\over 2^{n/2}}\,
             p_{\rm F}^n &; $n$ even, \cr
             \phantom{ \bigg[ } 0 &; $n$ odd. \cr}
\end{equation}
Comparing (\ref{antoy}) with (\ref{a2a3}) for $n=2$, one obtains the
relation $-\lambda_1=\frac{3}{2}\,p_{\rm F}^2$ between the low-energy
parameter $\lambda_1$ and the Fermi momentum \cite{shape}. Finally,
it is possible to calculate the leading term in the spectral function
$s(y,\rho)$ in (\ref{sresult}). The result is again a Gaussian
distribution:
\begin{equation}
   s_{\rm toy}(y,\rho) = {1\over\sqrt{2\pi}\,\sigma_y}\,
   \exp\bigg\{-{(y-1+\rho)^2\over 2\sigma_y^2}\bigg\} \,.
\end{equation}
The width $\sigma_y$ has been defined in (\ref{M1M2}). Such a
Gaussian model for the spectral function, supplemented by a radiative
tail extending to smaller values of $y$, has been used in the
analysis of Ali and Greub \cite{Greub}.

At this point, we have to stress that our toy model is presented for
pedagogical purposes only; we do not claim that it provides a
realistic description of the spectral function. In fact, this simple
model of the ``Fermi motion'' is inconsistent with QCD. Note that the
vanishing of the odd moments of the distribution function, which
implies the symmetry of the spectral function around $y=1-\rho$, is a
consequence of rotational invariance. As such, it is unavoidable in a
model where the time-component of the heavy quark momentum is
neglected. In QCD, however, there is no reason why any of the
coefficients $a_n$ (except $a_1$) should vanish. Hence, we expect
that the physical spectral function is asymmetric. In the following
section, we shall discuss in more detail the correct generalization
of the toy model in the context of QCD. This will lead us to the
concept of a universal light-cone structure function, which replaces
the distribution function $\phi(|\vec p_b|)$.

\section{The Light-Cone Structure Function}
\label{sec:5}

The alert reader will have realized the close analogy of our
discussion in Sect.~\ref{sec:3} with deep inelastic scattering. In
this section, we will exploit this relationship. From now on, we will
neglect the mass of the strange quark and set $\rho=0$. We expect
this to be an excellent approximation. For instance, the coefficients
$a_2(\rho)$ and $a_3(\rho)$ in (\ref{a2a3}) change by less than 1.5\%
when $m_s$ is varied between 0 and 0.4 GeV.

For $\rho=0$, the vector
\begin{equation}
   n_\mu = 2 (v-\hat p)_\mu \Big|_{y=1}
\end{equation}
is a null vector on the forward light cone satisfying $n^2=0$ and
$n\cdot v=1$. Let us denote the scalar product of a four-vector $p$
with $n$ by $n\cdot p\equiv p_+$. In the rest frame of the $B$-meson,
we are free to choose $n_\mu=(1,0,0,1)$, such that $p_+=p^0+p^3$.
Moreover, we can simplify expressions by using the light-cone gauge
(LCG) $n\cdot A=0$. From (\ref{anres}) and (\ref{a2a3}), it then
follows that
\begin{equation}
   a_n(0) = A_n = \langle\,(i D_+)^n\,\rangle
   \stackrel{\rm LCG}{\phantom{|}=\phantom{|}}
   \langle\,(i\partial_+)^n\,\rangle \,.
\end{equation}
This is the correct generalization of (\ref{antoy}). Note that
$i\partial_+$ is the operator corresponding to the light-cone
residual momentum $k_+$ of the $b$-quark in the $B$-meson.

Using this notation, the spectral function $s(y)\equiv s(y,0)$ in
(\ref{sresult}) takes the form
\begin{eqnarray}\label{convol}
   s(y) &=& \bigg\langle \delta\bigg( 1-y + {i D_+\over m_b} \bigg)
    \bigg\rangle +~\hbox{less singular terms} \nonumber\\
   &=& \int{\rm d}k_+\,\delta\bigg( 1-y + {k_+\over m_b} \bigg)\,
    \Big\{ f(k_+) + {\cal{O}}(1/m_b) \Big\} \,,
\end{eqnarray}
where
\begin{equation}\label{fdef}
   f(k_+) = \langle\,\delta(i D_+ - k_+)\,\rangle
   = {\langle B(v)|\,\bar h_v\,\delta(i D_+ - k_+)\,h_v\,
   |B(v)\rangle \over \langle B(v)|\,\bar h_v\,h_v\,|B(v)\rangle}
\end{equation}
is a universal {\sl structure function}, which determines the
probability to find a $b$-quark with light-cone residual momentum
$k_+$ inside the $B$-meson. Since this function is defined in terms
of a matrix element in the heavy quark effective theory, it is
independent of the $b$-quark mass. The moments of $f(k_+)$ are given
directly in terms of the hadronic matrix elements $A_n$ defined in
(\ref{Andef}),
\begin{equation}
   A_n = \int{\rm d}k_+\,k_+^n\,f(k_+) \,.
\end{equation}
The corresponding moment expansion reads
\begin{equation}
   f(k_+) = \sum_{n=0}^\infty\,{(-1)^n\over n!}\,A_n\,
   \delta^{(n)}(k_+) = \delta(k_+) - {\lambda_1\over 6}\,
   \delta''(k_+) - {A_3\over 6}\,\delta'''(k_+) + \ldots \,.
\end{equation}

Equation~(\ref{convol}) shows very nicely the structure of the
$1/m_b$ expansion proposed in this paper: The photon spectrum is a
convolution of the free quark decay spectrum (i.e.\ a
$\delta$-function) with a nonperturbative distribution function. The
leading term in the $1/m_b$ expansion of this function is given by
the universal structure function $f(k_+)$. This is the correct
generalization of the toy model discussed in the previous section.
The important distinction is that the light-cone momentum $k_+$
contains the time-component of the residual momentum $k$. Thus, the
odd moments of the structure function are not forced to vanish by
rotational invariance, and $f(k_+)$ is in general not symmetric
around $k_+=0$.

One can consider (\ref{convol}) as the recipe for a systematic and
consistent implementation of the leading bound-state corrections,
even when one goes beyond the leading order in perturbation theory.
Needless to say, however, the structure function $f(k_+)$ cannot be
calculated from first principles. One option is to extract this
universal function from experimental data. The photon spectrum in
rare $B\to X_s\,\gamma$ decays is an ideal place for this.
Alternatively, one may try to obtain a QCD-based prediction for
$f(k_+)$ using nonperturbative techniques such as lattice gauge
theory or QCD sum rules. For this purpose, it may be useful to relate
the structure function to a forward matrix element of a
gauge-invariant, bilocal operator. Let us define the Fourier
transform of $f(k_+)$ as
\begin{equation}
   f(t) = \int{\rm d}k_+\,f(k_+)\,e^{-i k_+ t} \,.
\end{equation}
It then follows that
\begin{equation}\label{Grozin}
   f(t) \equiv {\langle B(v)|\,\bar h_v(0)\,{\rm P}\,
   \exp\bigg( -i\int\limits_0^z\!{\rm d}x_\mu A^\mu(x)\bigg)
   \,h_v(z)\,|B(v)\rangle \over
   \langle B(v)|\,\bar h_v(0)\,h_v(0)\,|B(v)\rangle} \,,
\end{equation}
where $z=t\,n$ is a four-vector on the light cone satisfying $z^2=0$
and $z\cdot v=t$, ``P'' denotes path ordering, and the integral is
along a straight line. In light-cone gauge, the phase factor equals
unity. The function $f(t)$ describes the spatial distribution of the
$b$-quark inside the $B$-meson.

Finally, we note that the residual momentum structure function
$f(k_+)$ obeys a simple relation to the usual structure function
$b_B(x)$, which determines the probability to find in the $B$-meson a
$b$-quark with total light-cone momentum fraction $x$. Using that
$p_b=m_b v+k$ and $n\cdot v=1$, we have
\begin{equation}
   x\equiv {(p_b)_+\over (p_B)_+} = {m_b+k_+\over m_B} \,,
\end{equation}
and hence
\begin{equation}
   b_B(x)\,{\rm d}x = \Big\{ f(k_+) + {\cal{O}}(1/m_b) \Big\}\,
   {\rm d}k_+ \,;\quad k_+ = m_B\,x - m_b \,.
\end{equation}
{}From the requirement that $0\le x\le 1$, it follows that the
allowed range for the light-cone residual momentum is $-m_b\le k_+\le
m_B-m_b$. In the limit $m_b\to\infty$, this becomes
\begin{equation}\label{krange}
   -\infty < k_+ \le \bar\Lambda \,,
\end{equation}
where $\bar\Lambda$ denotes the asymptotic value of the mass
difference $m_B-m_b$ and can be identified with the effective mass of
the light degrees of freedom in the $B$-meson \cite{AMM,SR3}. An
interesting consequence of (\ref{krange}) is that it determines the
kinematic endpoint for the photon energy in $B\to X_s\,\gamma$
decays. From (\ref{convol}), we find that
\begin{equation}
   y_{\rm max} = 1 + {\bar\Lambda\over m_b} + {\cal{O}}(m_b^{-2})
   = {m_B\over m_b} + {\cal{O}}(m_b^{-2}) \,.
\end{equation}
This is in fact consistent with the physical endpoint given in
(\ref{kine}).

Knowing the moments of $f(k_+)$, it is straightforward to calculate
the moments of the structure function $b_B(x)$. We find
\begin{equation}
   \int\limits_0^1\!{\rm d}x\,(1-x)^n\,b_B(x)
   = {1\over m_B^n}\,\sum_{k=0}^n\,(-1)^k\,
   \bigg(\begin{array}{c} n\\ k\end{array}\bigg)\,
   A_k\,\bar\Lambda^{n-k} + {\cal{O}}(m_B^{-n-1}) \,.
\end{equation}
In particular, this leads to the sum rules
\begin{eqnarray}
   \int\limits_0^1\!{\rm d}x\,b_B(x) &=& 1 \,, \nonumber\\
   \int\limits_0^1\!{\rm d}x\,(1-x)\,b_B(x) &=&
    {\bar\Lambda\over m_B} + {\cal{O}}(m_B^{-2})\,, \nonumber\\
   \int\limits_0^1\!{\rm d}x\,(1-x)^2\,b_B(x) &=&
   {1\over m_B^2}\,\bigg(\bar\Lambda^2 - {\lambda_1\over 3}\bigg)
     + {\cal{O}}(m_B^{-3}) \,.
\end{eqnarray}
The second relation has been derived previously, in a different
context, in Refs.~\cite{Riec,Burk}. Note that it implies the lower
bound $\bar\Lambda>0$, which in view of the recent criticism
\cite{BiUr} of the Guralnik-Manohar bound $\bar\Lambda>237$ MeV
\cite{GuMa} seems less trivial than one may think.

\section{Relation to $B\to X_u\,\ell\,\bar\nu$ Decays}
\label{sec:6}

In this section, we discuss an interesting relation between the
leading nonperturbative corrections in rare and semileptonic
inclusive $B$-decays. This connection may help to obtain a
model-independent determination of the element $V_{ub}$ of the
Kobayashi-Maskawa matrix. In the limit $m_u=m_\ell=0$, the inclusive
lepton spectrum in $B\to X_u\,\ell\,\bar\nu$ decays can be written as
\begin{equation}
   {{\rm d}\Gamma\over{\rm d}y}
   = {G_F^2\,|\,V_{ub}|^2\over 96\pi^3}\,m_b^5\,
   \Big[ F(y)\,\Theta(1-y) + F(1)\,S(y) \Big] \,;\qquad
   y = {2 E_\ell\over m_b} \,.
\end{equation}
In this case, the kinematic variable $y$ denotes the rescaled lepton
energy, and $F(y)$ is a slowly varying function of $y$. Apart from
small nonperturbative corrections of order $1/m_b^2$
\cite{Bigi,MaWe}, it is given by parton model kinematics: $F(y)\simeq
y^2(3-2y)$. Close to the endpoint, one can replace $F(y)\simeq
F(1)\simeq 1$, up to corrections of order $1/m_b^2$. The shape
function $S(y)$, on the other hand, is a rapidly varying, genuinely
nonperturbative object \cite{shape}. It is non-zero only in a small
region around the endpoint of the spectrum. Using the resummation
technique developed in Sect.~\ref{sec:3}, we find that
\begin{eqnarray}\label{ThetS}
   \Theta(1-y) + S(y) &=& \bigg\langle
    \Theta\bigg( 1 - y + {i n\cdot D\over m_b} \bigg) \bigg\rangle
    +~\hbox{less singular terms} \nonumber\\
   &=& \int{\rm d}k_+\,\Theta\bigg( 1-y + {k_+\over m_b} \bigg)\,
    \Big\{ f(k_+) + {\cal{O}}(1/m_b) \Big\} \,.
\end{eqnarray}
Here, $n=2(v-p_\ell/m_b)$ is again a null vector when $y=1$, and
hence $f(k_+)$ coincides with the universal function defined in
(\ref{fdef}). We observe that, close to the endpoint region, the
lepton spectrum in semileptonic $B\to X_u\,\ell\,\bar\nu$ decays can
again be written as a convolution of the free quark decay
distribution (i.e.\ a step function) with the structure function
$f(k_+)$. Comparing the above relation with (\ref{convol}), we obtain
\begin{equation}
   s(y) = -{\partial\over\partial y}\,\Big[ \Theta(1-y) + S(y)
   \Big] +~\hbox{less singular terms.}
\end{equation}
Let us integrate this equation to obtain
\begin{eqnarray}\label{sSrela}
   \Theta(1-y) + S(y) &\simeq& \int\limits_y^\infty\!{\rm d}y'\,
    s(y') \,, \nonumber\\
   \int\limits_y^\infty\!{\rm d}y'\,\Big[
   \Theta(1-y') + S(y') \Big] &\simeq& \int\limits_y^\infty\!
   {\rm d}y'\,(y'-y)\,s(y') \,.
\end{eqnarray}
These relations are exact up to corrections of order $1/m_b$ or
$\alpha_s$.

The main goal of the study of $B\to X_u\,\ell\,\bar\nu$ decays is to
extract the element $V_{ub}$ of the Kobayashi-Maskawa matrix. The
experimental analysis is very complicated, as there is only a tiny
window close to the endpoint region where the signal is not
overshadowed by the large background from $B$-decays into charmed
particles. What can be measured is an integral over the endpoint
region,
\begin{equation}
   \widehat\Gamma_u(E_0) \equiv \int\limits_{E_0}^\infty\!
   {\rm d}E_\ell\,
   {{\rm d}\Gamma(B\to X_u\,\ell\,\bar\nu)\over{\rm d}E_\ell} \,,
\end{equation}
where $E_0$ is above the kinematic endpoint for $B\to D\,\ell\,
\bar\nu$ transitions, i.e.\ $E_0>2.3$ GeV. The traditional way to
extract a value of $V_{ub}$ from such a measurement is to compare
$\widehat\Gamma_u(E_0)$ with the predictions of various quark models.
Since the endpoint region of the lepton spectrum is strongly affected
by nonperturbative effects, this procedure suffers from a
considerable amount of model-dependence. Currently, the value of
$V_{ub}$ obtained following this strategy has a theoretical
uncertainty of at least a factor 2 \cite{Vub1,Vub2}.

Based on the results obtained in this paper and in Ref.~\cite{shape},
we propose a new strategy to extract $V_{ub}$ with little
model-dependence. The idea is to use the second equation in
(\ref{sSrela}) to relate the integral $\widehat\Gamma_u(E_0)$ to a
weighted integral over the photon spectrum in rare decays. Defining
\begin{equation}\label{GamsE0}
   \widehat\Gamma_s(E_0) \equiv {2\over m_B}
   \int\limits_{E_0}^\infty\!{\rm d}E_\gamma\,(E_\gamma-E_0)\,
   {{\rm d}\Gamma(B\to X_s\,\gamma)\over{\rm d}E_\gamma} \,,
\end{equation}
and using that $|\,V_{tb} V_{ts}^*|\simeq |\,V_{cb}\,|$, we find the
remarkable relation
\begin{equation}\label{magic}
   \bigg| {V_{ub}\over V_{cb}} \bigg|^2 \simeq
   \bigg| {V_{ub}\over V_{tb} V_{ts}^*} \bigg|^2
   = {3\alpha\over\pi}\,|c_7(m_b)|^2\,\eta_{\rm QCD}\,
   {\widehat\Gamma_u(E_0)\over\widehat\Gamma_s(E_0)}
   + {\cal{O}}\bigg( {\Lambda_{\rm QCD}\over m_b}\bigg) \,,
\end{equation}
where $\eta_{\rm QCD}$ contains radiative corrections, which have so
far been neglected in this paper. The Wilson coefficient $c_7(m_b)$
can be calculated in perturbation theory. In the standard model, it
is known to next-to-leading logarithmic accuracy \cite{Misi,Ciuc}.
Hence, from a measurement of the integrated quantities
$\widehat\Gamma_u(E_0)$ and $\widehat\Gamma_s(E_0)$ one obtains a
direct determination of $|V_{ub}/V_{cb}|$. The fact that the
right-hand side in (\ref{magic}) must be independent of $E_0$
provides a constraint, which can help in the analysis of the data. By
taking the ratio of the integrated decay rates, we are able to reduce
hadronic uncertainties to the level of power corrections, the leading
ones being of order $\Lambda/m_b\simeq 0.1$, where we take
$\Lambda\simeq 500$ MeV as a typical low-energy scale of the strong
interactions. It is thus not inconceivable that the theoretical
uncertainties in (\ref{magic}) can be controlled to a level of, say,
10--30\%, corresponding to an uncertainty in $|V_{ub}/V_{cb}|$ of
5--15\%. This would be a major improvement over the present
situation.

To achieve such an accuracy, it is necessary to study in detail the
QCD correction factor $\eta_{\rm QCD}$ in (\ref{magic}), which arises
when radiative corrections are included in the operator product
expansion of the inclusive decay rates. We shall briefly discuss the
qualitative structure of these corrections; a more complete treatment
will be presented elsewhere \cite{radi}. In the free quark decay
model, the one-loop radiative corrections to $B\to X_s\,\gamma$ and
$B\to X_u\,\ell\,\bar\nu$ decays have been investigated by several
authors \cite{Ali,AlPi,CCM,JeKu}. For the integrated quantities
$\widehat\Gamma_i(E_0)$, they have the general structure ($i=u$ or
$s$)
\begin{eqnarray}\label{Gamipart}
   \widehat\Gamma_i^{\rm parton}(E_0) &\propto& (1-y_0)\,
    \Theta(1-y_0)\,\bigg\{ 1 - {2\alpha_s\over 3\pi}\,\bigg[
    \ln^2(1-y_0) + a_i\,\ln(1-y_0) + b_i \bigg] \bigg\} \nonumber\\
   &&\mbox{}+ {\cal{O}}\Big[ (1-y_0)^2 \Big] \,,
\end{eqnarray}
where $y_0=2 E_0/m_b$. The Sudakov-type double logarithms are
universal and enter both quantities with the same coefficient. This
statement is true to all orders in perturbation theory
\cite{Suda1,Suda2}. Note that the result for $\widehat\Gamma_s(y_0)$
is still of this form when one takes into account the finite strange
quark mass. This effect is known to modify the endpoint behavior of
the spectrum in the very endpoint region $1-y_0\sim\rho={\cal{O}}
(1/m_b^2)$ \cite{Ali}. However, since the integral in (\ref{GamsE0})
extends over a larger region of order $1/m_b$, the corrections
induced by $\rho\ne 0$ are subleading. They are of the same magnitude
as terms of order $(1-y_0)^2$, which we neglect in (\ref{Gamipart}).

Since integrations over $y$ and $k_+$ commute, we can incorporate
bound-state corrections by convoluting the parton model result
(\ref{Gamipart}) with the structure function $f(k_+)$, as we did for
the tree-level expressions in (\ref{convol}) and (\ref{ThetS}). When
we then take the ratio of $\widehat\Gamma_u(E_0)$ and
$\widehat\Gamma_s(E_0)$, the large double-logarithmic corrections
cancel. We find
\begin{equation}\label{etaQCD}
   \eta_{\rm QCD} = 1 - {2\alpha_s\over 3\pi}\,\Big[
   (a_u-a_s)\,\ln r + (b_u-b_s) \Big] \,.
\end{equation}
Here
\begin{equation}
   \ln r = { \displaystyle
       \int\limits_{m_b(y_0-1)}^{\bar\Lambda}\!\!\!{\rm d}k_+\,
       f(k_+)\,\bigg( 1-y_0+{k_+\over m_b} \bigg)\,
       \ln\bigg( 1-y_0+{k_+\over m_b} \bigg) \over
       \displaystyle
       \int\limits_{m_b(y_0-1)}^{\bar\Lambda}\!\!\!{\rm d}k_+\,
       f(k_+)\,\bigg( 1-y_0+{k_+\over m_b} \bigg) }
\end{equation}
is a nonperturbative parameter of order $\Lambda_{\rm QCD}/m_b$.
Using the fact that the logarithm is a monotonic function over the
range of integration, and that $1-y_0+\bar\Lambda/m_b=1-2 E_0/m_B$ up
to corrections of order $1/m_b^2$, we find the bound
\begin{equation}
   - \ln r > - \ln\bigg( 1 - {2 E_0\over m_B} \bigg) \,.
\end{equation}
A precise calculation of $\ln r$, however, requires some knowledge of
the nonperturbative structure function $f(k_+)$. The coefficients
$a_s$ and $b_s$ in (\ref{etaQCD}) can be extracted from the analysis
of Ali and Greub \cite{Ali}. We obtain $a_s=3/2$ and $b_s=2\pi^2/3 +
1$. In $b_s$, we have neglected small contributions from operators
other than $O_7$ in the short-distance expansion of the effective
Hamiltonian (\ref{Heff}). Unfortunately, the existing analytical
calculations of the radiative corrections to the lepton spectrum in
$B\to X_u\,\ell\,\bar\nu$ decays disagree on the value of $b_u$. We
find $a_u=19/6$ and $b_u=\pi^2-23/12+\delta$, where $\delta=0$
according to Jezabek and K\"uhn \cite{JeKu}, while $\delta\simeq 3/8$
according to Corb\`o \cite{CCM}. For the correction factor $\eta_{\rm
QCD}$, we find
\begin{eqnarray}
   \eta_{\rm QCD} &=& 1 - {2\alpha_s\over 9\pi}\,\bigg( 5\ln r
    + \pi^2 + \delta - {35\over 4} \bigg) \nonumber\\
   &\simeq& 1 + \bigg[ 1.54 - 1.11\,\ln\bigg({r\over 0.2}\bigg)
    - 0.22\,\delta \bigg]\,{\alpha_s\over\pi} \,.
\end{eqnarray}
Since the controversial quantity $\delta$ enters with a small
coefficient, the corresponding uncertainty in $\eta_{\rm QCD}$ is
small. Using $\alpha_s/\pi\simeq 0.08$ and $r\simeq 0.2$, we expect
$\eta_{\rm QCD}\simeq 1.12$. We conclude that the perturbative
corrections are not unexpectedly large, but they deserve further
investigation. In particular, the nature of the single-logarithmic
corrections should be clarified. It is tempting to interpret the
$\ln r$-term in (\ref{etaQCD}) as a renormalization-group logarithm
arising from the scaling from large scales $\mu^2\simeq m_b^2$ down
to smaller scales $\mu^2\simeq m_b^2(1-y_0)\sim m_b\,\Lambda_{\rm
QCD}$, which are characteristic of the endpoint region. If this
interpretation is correct, it should be possible to resum these
logarithms using renormalization-group techniques. At present,
however, we cannot prove this assertion.

Let us finally compare our new strategy to alternative
model-independent determinations of $V_{ub}$ from exclusive
$B$-decays. Using heavy quark flavor symmetry, one can in principle
extract $V_{ub}$ from a comparison of exclusive decays $B\to
h\,\ell\,\bar\nu$ and $D\to h\,\ell\,\bar\nu$, where $h$ is a light
hadron such as $\pi$ or $\rho$ (see, e.g., Refs.~\cite{hl1,hl2,hl3}).
The problem is that the comparison must be done close to the zero
recoil limit. Thus, one is restricted to a small fraction of
phase-space. Presently, there is no convincing experimental evidence
for exclusive charmless $B$-decays. But even if such an analysis
becomes feasible as high-statistics data from a $B$-factory become
available, the theoretical uncertainties will be of order $1/m_c$
instead of $1/m_b$. We thus believe that our new approach, based on a
comparison of inclusive decay spectra, is preferable from the point
of view of both theoretical uncertainty and experimental feasibility.

\section{Summary and Conclusions}
\label{sec:7}

We have presented a systematic, QCD-based analysis of bound-state
corrections to the photon spectrum in inclusive $B\to X_s\,\gamma$
decays. Using the operator product expansion and the heavy quark
effective theory, we are able to resum the leading nonperturbative
effects into a universal structure function $f(k_+)$, which describes
the distribution of the light-cone residual momentum of the $b$-quark
inside the $B$-meson. This formalism provides the generalization of
the phenomenological concept of the ``Fermi motion'' in the context
of QCD. We find that the moments of the structure function are given
by a set of universal forward matrix elements of higher-dimension
operators. The characteristic width of the photon spectrum is related
to the expectation value of the kinetic energy of the heavy quark
inside the $B$-meson. As a by-product, we obtain the bound
$\lambda_1<0$ for one of the parameters of the heavy quark effective
theory.

The formalism presented here is rather general. So far, it has been
applied to the photon spectrum in rare decays and to the lepton
spectrum in $B\to X_u\,\ell\,\bar\nu$ transitions \cite{shape}.
Applications to other processes such as $B\to X_c\,\ell\,\bar\nu$
\cite{future} or purely hadronic decays are possible, too. The fact
that the leading nonperturbative corrections to inclusive decay
spectra can be traced back to a universal structure function leads to
interesting relations between different processes. In
Sect.~\ref{sec:6}, we have shown that a weighted integral over the
photon spectrum in $B\to X_s\,\gamma$ decays is related to an
integral over the endpoint region of the lepton spectrum in $B\to
X_u\,\ell\,\bar\nu$ decays. Based on this connection, we have
proposed a model-independent way to extract the ratio
$|V_{ub}/V_{cb}|$ of elements of the Kobayashi-Maskawa matrix.
Hadronic uncertainties enter this determination only at the level of
$1/m_b$ corrections. We estimate that using this method one could
extract $V_{ub}$ with a theoretical uncertainty of about 20\%, which
is an order of magnitude better than the present theoretical
uncertainty in this parameter.

Since the $b$-quark structure function is a rather fundamental
object, one should try to calculate it using nonperturbative
techniques such as lattice gauge theory or QCD sum rules. We believe
that the relations derived in Sect.~\ref{sec:4} will be helpful for
such calculations. In particular, it may be of advantage to calculate
the Fourier transform of the structure function, which is given by
the forward matrix element of a bilocal operator. Any theoretical
insight into the behavior of the structure function and its moments
will have a direct impact on the analysis of inclusive decay spectra,
from which one hopes to extract accurate values for some standard
model parameters or, in the case of rare decays, even indications for
new physics beyond the standard model. In this context, we emphasize
that the phenomenological model of Altarelli et al.~\cite{ACM}, which
has been widely used to account for bound-state corrections,
corresponds to a simple Gaussian model for the structure function and
is not entirely consistent with QCD. A measurement of some of the
moments of the shape function in semileptonic decays \cite{shape}, or
of the photon spectrum in rare decays, could help to find deviations
from this model. More generally, such a measurement would provide us
with some fundamental QCD matrix elements and is thus interesting in
its own right.

Finally, we mention that before the formalism developed here can be
applied to the analysis of data, it is necessary to include QCD
radiative corrections. This can be done by calculating virtual and
real gluon corrections in the free quark decay model, and convoluting
the result with the structure function $f(k_+)$. This procedure has
its subtleties, however, due to the presence of endpoint
singularities in the perturbative expansion. We have indicated this
for the important case of the ratio of integrated decay rates in
(\ref{magic}). A more complete discussion will be given elsewhere
\cite{radi}.

\bigskip\bigskip
{\it Acknowledgements:\/}
It is a pleasure to thank Ahmed Ali, Christoph Greub, Andrey Grozin,
Matthias Jamin, Thomas Mannel, and Daniel Wyler for inspiring
discussions. I am particularly indebted to Andrey for pointing out
relation (\ref{Grozin}).

\end{document}